\documentclass[preprint]{aastex}
\usepackage{lscape,graphicx}

\slugcomment{To appear in the Astronomical Journal}

\shorttitle{The INT Search for Metal-Poor Stars}
\shortauthors{Allende Prieto et al.}

\begin{document}


\title{The INT Search for Metal-Poor Stars. Spectroscopic Observations and
Classification via Artificial Neural Networks\footnote{Based on observations made with INT operated on the
island of La Palma by the Isaac Newton Group in the Spanish
Observatorio del Roque de los Muchachos of the Instituto de Astrofisica de
Canarias.}}


\author{Carlos Allende Prieto\altaffilmark{1}, Rafael Rebolo\altaffilmark{2}, Ram\'on J. Garc\'{\i}a L\'opez\altaffilmark{3}, Miguel Serra-Ricart}
\affil{Instituto de Astrof\'{\i}sica de Canarias, \\ E-38200, La Laguna,
Tenerife,   SPAIN}

\author{Timothy C. Beers}
\affil{Department of Physics and Astronomy, Michigan State University, \\ East
Lansing, MI 48824, USA}

\author{Silvia Rossi}
\affil{Instituto Astron\^omico e Geof\'{\i}sico, Universidade de S\~ao Paulo, \\
Av. Miguel Stefano 4200, 04301-908, S\~ao Paulo SP, BRAZIL}

\author{Piercarlo Bonifacio and Paolo Molaro}
\affil{Osservatorio Astronomico di Trieste, \\ Via Tiepolo 11, I-34131,
Trieste, ITALY}

\altaffiltext{1}{Present address: McDonald Observatory
and Department of Astronomy, The University of Texas at Austin, RLM 15.308,
Austin, TX 78712-1083, USA}
\altaffiltext{2}{Also Consejo Superior de Investigaciones Cient\'{\i}ficas (CSIC). SPAIN}
\altaffiltext{3}{Also Departamento de Astrof\'\i sica, Universidad de La Laguna,
Avd. Astrof\'\i sico Francisco S\'anchez s/n, E-38071, La Laguna,   Tenerife, SPAIN}


\begin{abstract}

With the dual aims of enlarging the list of extremely metal-poor stars
identified in the Galaxy, and boosting the numbers of moderately
metal-deficient stars in directions that sample the rotational properties of
the thick disk, we have used the 2.5m Isaac Newton Telescope and the
Intermediate Dispersion Spectrograph to carry out a survey of brighter
(primarily northern hemisphere) metal-poor candidates selected from the HK
objective-prism/interference-filter survey of Beers and collaborators.  Over
the course of only three observing runs (15 nights) we have obtained
medium-resolution ($\lambda/\delta\lambda \simeq 2000$) spectra for 1203
objects ($V \simeq 11-15$).  Spectral absorption-line indices and radial
velocities have been measured for all of the candidates.  Metallicities,
quantified by [Fe/H], and intrinsic $(B-V)_0$ colors have been estimated for
731 stars with effective temperatures cooler than roughly 6500 K, making use of
artificial neural networks (ANNs), trained with spectral indices.  We show that this method performs
as well as a previously explored Ca {\sc II} K calibration technique, yet it
presents some practical advantages.  Among the candidates in our sample, we
identify 195 stars with [Fe/H] $\le -1.0$, 67 stars with [Fe/H] $\le -2.0$, and
12  new stars with [Fe/H] $\le -3.0$.  Although the {\it effective yield} of
metal-poor stars in our sample is not as large as previous HK survey follow-up
programs, the {\it rate} of discovery per unit of telescope time is quite high.

Further development of the ANN technique, with the networks being fed the
entire spectrum, rather than just the spectral indices, holds the promise to
produce fast, accurate, multi-dimensional spectral classifications (with the
associated physical parameter estimates), as is required to process the large
data flow provided by present and future instrumentation.

\end{abstract}




\section{Introduction}
\label{sec1}

High-resolution, high-signal-to-noise stellar spectroscopy provides a wealth of
information on the physics and chemistry of stellar atmospheres, as well as on
the  dynamical and chemical evolution of our Galaxy.  In the past, however, the
small number of targets has limited the depth of information useful for
constraining several important problems of the early history of the Galaxy.
For example, measurements of chemical abundances and stellar motions of a
sample  of thousands of metal-deficient stars, ideally selected without
kinematic bias,  is required to address the most important problems relating to
the kinematic structure of the stellar populations of the Milky Way (see, e.g.,
Beers et al. 2000; Chiba \& Beers 2000).  Similarly, the identification of
trends, and in particular, the measurement of spreads in the relative
abundances of individual elements provides required constraints on the nature
of what was most likely a very inhomogeneous early evolution history 
(Ryan, Norris, \& Beers 1996; Tsujimoto \& Shigeyama 1998; Argast et al.
2000).  To achieve these goals, large collaborative efforts are already in
progress (see Beers 1999 for a summary).

Here we report on medium-resolution ($\sim 2$ \AA\ ) spectroscopic observations
for 731 stars obtained at the Isaac Newton Telescope (INT) on La Palma.  The
remaining $\sim 400$ stars with spectroscopy obtained during the INT follow-up
observations turned out to be generally hotter than the main-sequence turnoff
of an old stellar population, and are dominated by field horizontal-branch
stars, blue metal-poor stars or blue stragglers, and other hot stars such as O-
and B-type subdwarfs.  These hot stars provide additional information on the
kinematics of the Galaxy (e.g., Sommer-Larsen et al. 1997; Wilhelm, Beers, \&
Gray 1999), and will be the subject of a future paper.

The stars under discussion were previously identified as metal-poor candidates
from the HK objective-prism/interference-filter survey of Beers and
collaborators (as described by Beers, Preston, \& Shectman 1985; Beers,
Preston, \& Shectman 1992, hereafter BPS).  The HK objective-prism technique
obtains wide-field (5 $\times$ 5 degree) photographic plates of  widened
spectra, with the interference filter selecting a $\sim 150$ \AA\  region
around the Ca~{\sc ii} H and K absorption lines, reaching limiting magnitudes
of $B \sim 15-15.5$.  Initially, the HK survey made use of the Curtis Schmidt
telescope at Cerro Tololo Inter-American Observatory for a survey of some 4000
square degrees of the southern-hemisphere sky.  Later, the survey was extended
to include some 3000 square degrees in the northern hemisphere, using the
Burrell Schmidt telescope at Kitt Peak National Observatory.  Visual inspection
with a low-power binocular telescope was originally carried out by Preston, and
later by Beers, to produce a list of $\sim$ 10,000 candidate metal-poor stars,
a subset of which was observed with the INT.  Other follow-up photometry and
spectroscopy campaigns have been conducted in the northern and southern
hemispheres; papers describing their results have either already been published
(Schuster et al.  1996; Norris, Ryan, \& Beers 1999; Anthony-Twarog et al.
2000) or are in preparation.

Beers et al. (1990) originally reported on a method for the estimation of
stellar metal abundance based on spectra of similar resolution to our present
program.  The Beers et al. procedure made use of synthetic spectra generated
from the Kurucz model atmospheres (Kurucz 1993) to derive a relationship
between the metallicity [Fe/H], the pseudo-equivalent width of the Ca~{\sc ii}
K line, and predicted broadband $(B-V)_0$ color, as obtained from the Revised
Yale Isochrones (Green 1988; King, Demarque, \& Green 1988).  This approach
employed interpolated polynomial fits in the Ca~{\sc ii} K {\it vs} $(B-V)_0$
plane, forcing agreement between the metallicities and colors provided by the
models, and those derived from high-resolution spectroscopy and broadband
photometry for a set of standard stars covering the range of metallicities and
colors of interest.  This method has recently been refined (Beers et al. 1999)
by making use of locally weighted multi-dimensional regressions instead of
polynomial fits, and by greatly expanding the number of standards used in the
calibration.  The 551 standard stars with accurate high-resolution metallicity
determinations and broadband photometry discussed by Beers et al.  (1999) are
used in the present paper to train and test the performance of the ANNs. No
overlap exists between the standard stars and the 731 program stars.

The outline of this paper is as follows.  In \S 2 we summarize the
spectroscopic observations and data reduction, as well as the measurement of
radial velocities and line indices for the stars.  In \S 3 we discuss the use
of the ANN technique for the estimation of stellar metallicity and broadband
colors for the INT sample, and compare with preliminary results derived from
application of the Beers et al. (1999) approach.  In \S 4 we discuss the
results of our search for low metallicity stars, and compare with the previous
searches of BPS.  Section 5 presents our conclusions.

\section{Observations, Reductions, and Measurement of Radial Velocities and
Line Indices}
\label{sec2}

Spectroscopic observations of our sample were carried out with the 2.5-m INT at
the Observatorio del Roque de los Muchachos, on the island of La Palma (Spain),
during three runs between 1995 March and 1996 June (see Table 1).  The 235 mm
camera and the 1200 groove mm$^{-1}$ grating of the Intermediate Dispersion
Spectrograph (IDS) provided a resolving power ${\lambda}/{\Delta\lambda}
\simeq$ 2000, and a spectral range of 1100 \AA\ centered at 4150 \AA, as  shown
in the sample spectrum in Figure 1.  The $1024 \times 1024$ TEK3 CCD camera
that was used is roughly 60\% efficient over the observed spectral range.  Only
256 pixels were retained in the spatial direction, reducing the readout time to
about 40 seconds while still providing  a sufficient  range for optimal sky
subtraction. The cosmetics on this chip were excellent, with a typical
pixel-to-pixel sensitivity variation at a level of 1\%.    Our target
signal-to-noise ratio was $S/N > 20$, and required exposure times, under
conditions of good seeing, varied between 1 and 30 minutes for stars in the
apparent magnitude range 11 $\le V \le$ 15.

Data reduction was performed in the standard way within the IRAF\footnote{IRAF
is distributed by the National Optical Astronomy Observatories, which are
operated by the Association of Universities for Research in Astronomy, Inc.,
under cooperative agreement with the National Science Foundation.} environment,
and consisted of bias correction, flatfielding, variance-weighted
spectral extraction, and wavelength calibration.

\subsection{Radial Velocity Determinations}

The IDS is attached to the INT's Cassegrain focus, resulting in displacements
of the individual spectra due to flexure during the course of the night that
might enlarge the errors in the measured radial velocities.  To reduce this
source of uncertainty, we obtained calibration-lamp spectra for groups of 5--10
stellar spectra of targets selected from the same HK survey plate, and which
were therefore located in the same region of the sky.  One calibration lamp
spectrum was obtained at the end of the exposure of the last star in the group,
before pointing to the next star, and another  was taken before starting on the
new group, with the telescope already pointing to the new target.  In this way
we collected two calibration spectra for every group, as well as spectra for
two or three radial-velocity standards per night.  The geocentric radial
velocity for each star was measured by two methods: averaging the spectral
shifts measured in prominent absorption lines, and by cross-correlating the
spectra with a set of artificial templates and standard stars (see Beers et al.
1999, and references therein, for a complete description).  Corrections for the
Earth's rotation and orbital motion were then applied to take the velocities to
the heliocentric system.  We estimate a final accuracy for the radial
velocities of about 10 km s$^{-1}$, as verified by comparison with the radial
velocity standards.

\subsection{Line Index Measurement}

The measured geocentric radial velocities were used to place the wavelength
scale of the spectra at zero rest velocity.  We then determined
pseudo-equivalent widths for a number of prominent spectral features, summing
the total counts  around the required line, and using a straight average of the
flux in nearby side-bands to estimate the continuum level.  In some cases,
several wavelength intervals were used, as  previous work has been shown  that
the optimal band size depends on line strength. Table 2 lists the bands
employed.  KP uses the indices K6, K12, and K18; HP2 uses the indices HD12 and
HD24 (see Beers et al. 1999 for more details).  KP is sensitive to the stellar
calcium abundance, which is assumed to track the overall metallicity (as
quantified by [Fe/H]\footnote{$[A/B] = \log \left(\frac{N_A}{N_B}\right) - \log
\left(\frac{N_A}{N_B}\right)_{\odot}$, where $N_X$ and is the number density of
nuclei of an element X.}), but it also depends on effective temperature, and
more weakly, on surface gravity.  Figure 2 illustrates the relationship between
an independent estimate of [Fe/H] from high-resolution spectroscopic analyses in
the literature, [Fe/H]$_{\rm lit}$, as a function of the KP index for the 551
standard stars of Beers et al. used in this study.  The observed scatter about
the simple linear fit arises primarily from real differences in the effective
temperatures of the stars in the sample.

There are well-documented differences between the solar elemental abundance
ratios and those found in the photospheres of most metal-poor stars. In
particular, the ratio of calcium-to-iron abundance is known to vary with
metallicity (see, e.g., McWilliam 1997). The procedure we have employed to
assign metallicity estimates remains valid, however, since the overabundance of
calcium with respect to iron increases monotonically and with little scatter towards lower iron
abundances  for the vast majority of stars, and the empirical metallicity
calibration described below takes this variation into account.  The critical
pair of indices is completed with the HP2 or the H8 index, which provide
estimates of the strengths of the Balmer lines H-$\delta$ or H-8, respectively.
The well-known temperature sensitivity of the hydrogen lines (see, e.g.,
Fuhrmann, Axer, \& Gehren  1993) is captured by these indices.  Figure 3
illustrates the relationship between another temperature-sensitive quantity,
the broadband $(B-V)_0$ color, with HP2, using the same set of standard stars
from Beers et al. (1999).  In the present application, we explored
supplementing these temperature-sensitive indices with two others, centered on
the Ca~{\sc i} $\lambda$ 4226 \AA\ (the CAP index) and the CH G-band at 4300
\AA\ (the GP index).  The additional spectral information allowed
us to improve estimates of the metallicities and colors for the program stars,
and might be used in the future to measure other quantities, such as carbon
abundance.

A number of white dwarfs, hot O- and B-type subdwarfs, other stars bluer than
$B-V \sim 0.3$, including numerous field horizontal-branch and other A-type
stars, as well as a number of emission-line objects, mostly M-dwarfs with
active photospheres, were mistaken for metal-poor stars in the (visual)
identification of candidates from the  HK objective-prism survey.  To exclude
such objects, only the stars having metallic-line indices in the
range of sensitivity to [Fe/H], and with estimated colors in the range $0.3 \le
B-V \le 1.2$ ($0.1 \le$ HP2 $\le 6.0$), were extracted from the full set of
spectra, and are the subject of this paper.

Column (1) of Table 3 lists the star names of our sample.  Columns (2) and (3)
list the (1950.0 epoch) coordinates, generally accurate to a few arcseconds.
The Galactic longitude and latitude of the stars are provided in columns (4)
and (5), respectively. Column (6) lists the measured heliocentric radial
velocities, in units of km s$^{-1}$.  Column (7) lists an estimate of the
Johnson $V$-band apparent magnitude, when available in the {\it Guide Star
Catalogue} (Lasker et al. 1989).  Column (8) is the estimated de-reddened
$(B-V)_0$ color, listed as $BV$ (described below), in order to distinguish this
estimate from a true photometric measurement.  Columns (9)--(13) list the
observed line indices for the program stars in units of \AA.  Column (14)
lists the derived estimate of [Fe/H], as described below. 

For a number of candidates, the presence of two or more objects in the field of
view caused difficulties for the identification of the HK survey target. In
many such situations, it was decided to put multiple candidates on the slit (by
rotating the spectrograph) and obtain spectra of all possible sources.  These
multiple cases have been labeled in Table 3 according to their approximate
orientation on the sky.

A subset of the plates from the HK survey overlapped with one another, hence
there are occasional instances of multiple identifications.  Table 4 lists the
names of the multiply identified stars that were included among the INT
program sample.  A search of the SIMBAD catalog within 1 arcmin of the
listed positions of our program stars revealed a small number of likely
previous identifications; these are listed in Table 5.

Figure 4 shows the positions of the program stars in Galactic coordinates, the
upper and lower panels corresponding to the northern and southern hemispheres,
respectively.

\section{Classification of the Stars}
\label{sec3}

Previous application of ANNs to the classification of astronomical spectra has
demonstrated the great utility of this new tool.  For example, ANNs have been
applied to low-dispersion {\it IUE} spectra by Vieira \& Ponz (1995), allowing
the determination of spectral class to an accuracy of $\sim $1.1 subclasses.
More recently, Bailer-Jones, Irwin \& von Hippel (1998) used ANNs to classify
spectra of resolution $\sim$ 3 \AA\  extracted from the Michigan Spectral
Survey (Houk 1994) with an accuracy of 1.09 spectral subclasses, and obtained
correct luminosity classes for over 95\% of dwarfs and giants in their test
sample.

An optimum procedure would make use of the entire measured spectrum of a given
star as an input for an ANN, and ideally, provide a classification method to
extract the effective temperature, surface gravity, and metallicity of each
star (see, e.g., Snider et al. 2000).  Unfortunately, the full implementation
of this approach was not available at the beginning of our analysis, thus we
decided to adopt a previous step, one making use of the measured spectral
indices described above.  This classification method, based on limited
information input to the ANN, has nevertheless enabled reasonably accurate
estimates to be made of the $(B-V)_0$ colors and [Fe/H] for the sample of
metal-poor stars observed at the INT, and is described below.  It is worthwhile
noting that Bailer-Jones (2000) has made use of {\it synthetic} spectra to show
that, when spectrophotometry is available, i.e., the absolute energy
distribution is measured, ANNs fed with the full spectrum behave quite well in
the face of degradations in the signal-to-noise ratio and the resolving power.

The Beers et al. (1999) calibration has been applied to provide preliminary
estimates of [Fe/H] for the sample of metal-poor stars observed at the INT, and
serves as an independent reference to compare with the metallicities derived
from the ANN approach.  This calibration is based on the combination of the KP
index, and an Auto-Correlation Function (ACF) of the stellar spectrum, which
produces abundance estimates having external errors on the order of 0.15--0.20
dex over nearly the complete range of metallicities expected in the stellar
populations of the Galaxy, $-4.0 \le {\rm [Fe/H]} \le 0.0$.  The ACF employed
by Beers et al. takes advantage of the fact that, when a stellar spectrum is
correlated with itself, the peak value of the resulting (non-trivial
overlapping) function depends on the strengths of various metallic features in
the spectrum ``beating'' against one another.  The greater the numbers of such
lines, and the greater their strengths, the larger the signal.  As Beers et al.
discuss, the ACF approach provides complimentary information to the KP index
methodology, and performs particularly well for cooler or more metal-rich stars
where the KP index approaches saturation.  

There are a number of iterative steps that are applied in order to arrive at a
final estimate of stellar metallicity using the Beers et al. techniques.  In
the case of the present application, we do not yet have measured ACFs for all
of the INT program stars, so we base our comparison spectroscopic abundance
estimates solely on the KP index.  The estimate we make use of, [Fe/H]$_{\rm
K3}$ (adopting the Beers et al.  nomenclature), is expected to be close to, but
not exactly the same as, the metallicity estimate that will be obtained once
the full set of ACFs is available. The largest differences are expected for the
stars with metallicities greater than [Fe/H] $ \sim -1.0$, where the KP index
begins to saturate, especially for the cooler stars.  An approximate correction
for this effect, even in the absence of ACF measurements, has been applied
before making our comparisons.  Beers et al.  also describe an independent
means of predicting broadband $(B-V)_0$ colors, based on the Balmer line
indices, which is used as a check on the ANN's assignment of colors.

\subsection{Artificial Neural Networks}
\label{annsect}

Artificial neural networks are computational systems whose structure is based
on that of the human brain and nervous system.  In the present application we
employ a feed-forward neural network, where ``neurons'' are grouped into layers
(Murtagh 1991). There is one input layer where the information comes in, one or
several hidden layers where the information is processed, and one layer that
yields the output of the network computations.  For instance, a 2-5-6-1 ANN
architecture is comprised of a network with two input neurons, one output
neuron, and two hidden layers with five and six neurons each.  The neural
network provides a mapping, $f$, from a set of inputs to a set of desired
outputs. The set of  ``input''--``desired output'' pairs forms the pattern
sample.  Depending on how the ``desired output'' is defined, learning will be
supervised (Serra-Ricart et al. 1996) or unsupervised (Serra-Ricart et al.
1993).  The mapping, $f$, is approximated by application of a specific learning
algorithm (in this case, a back-propagation algorithm -- see Rumelhart, Hinton,
\& Williams 1986) to adjust the weights of connections from examples contained
in the pattern sample.

In the present case the pattern sample was the set of line indices listed in
Table 2  for a randomly chosen subset of 50\% of the standard stars described
in Beers et al.  (1999).  The other half of the sample is set aside to carry
out an independent test of the accuracy of the ANN. The list of standards
includes 551 stars with high-resolution determinations of [Fe/H], and
previously measured $(B-V)_0$ colors, which had medium-resolution spectroscopy
obtained with nine different telescope/spectrograph combinations at essentially
the same resolution as the INT follow-up spectra, over the course of the HK
survey.  The heterogeneity of the sample does not necessarily introduce
significant additional random error, as it was checked that variations of the
KP and HP/H8 indices measured at different observatories were at the level of 5
\% and 10 \%, respectively, roughly the expected measurement error for a single
spectrum with our minimum signal-to-noise ratio ($S/N = 20$).

The two primary difficulties generally encountered with presently studied ANN
classifiers are:

\begin{enumerate}

\item  Poor mapping from input examples of the $f$ function. In Figure 5 the
standard stars are plotted (with open circles) in the KP--HP2 plane together
with the INT program stars (filled circles), demonstrating that the standards
essentially map the input parameter space. It is possible to distinguish some
stellar classes in the KP--HP2 diagram, as can be observed in Figure 6.

\item  The ANN becomes ``over-trained.''  This means that the ANN has fully
matched the input data.  A large ANN can learn the entire pattern sample,
essentially memorizing the complete set of input examples instead of providing
a valid approximation of the mapping function $f$. To prevent over-training we
randomly separate the pattern sample into two groups of almost the same size,
then use one of them for the training task (the {\it training} sample), and the
other to test for convergence of the training task (the {\it testing} sample).

\end{enumerate}

Some of the spectral features quantified by the line indices were not measurable
in all the stars, for example, the HP2 index for the cooler stars.  Hence we
considered a number of different ANNs, depending on the available indices.
This scheme was also adopted in order to identify the ANN with the highest
performance for a given set of available indices and output parameters.  We
trained and ran 49 different configurations with some or all of the input
indices  described in Table 2 (KP, HP2, H8, CAP, and GP) to extract some or all
of the following parameters: the estimated $(B-V)_0$ color index, [Fe/H],
surface gravity, and the related parameter, luminosity.  The experiments showed
that the ANN was able to recover estimates of $(B-V)_0$  and [Fe/H] with
reasonable accuracy, but that surface gravity and luminosity remain a
challenge.  The ANN configurations that lacked at least the KP and one of the
two hydrogen indices (HP2 and H8) were shown to be unsatisfactory and were
discarded.  Finally, two different configurations were used to provide
estimates of the $(B-V)_0$ color index, and three different configurations were
used to provide estimates of metallicity.  Most of the stars in the INT sample
were classified by the optimal ANN configurations (\#4 and \#11), although the
differences with the other configurations considered, listed in Table 6, were
not large.

Figure 7 shows the results obtained for the {\it testing} sample of standard
stars after training with ANN configuration \#11.  The errors in metallicity
exhibit a Gaussian distribution, and do not depend significantly on
[Fe/H].  Likewise, Figure 8 shows that the ANN estimate of
$(B-V)_0$, $BV$, correlates well with the measured broadband color.  Note,
however, that the errors on estimates of color are much smaller for the {\it
hotter} (smaller $BV$) stars than for the {\it cooler} (larger $BV$) stars, for
which the error distribution deviates from a Gaussian distribution.  This is a
natural consequence of the Balmer lines' weakening towards lower effective
temperatures, and the resulting increase in the errors of their line index
measurements.  In order to check for possible correlations between errors in
[Fe/H] and  $BV$ we made use of one of the ANN configurations with two output
neurons (config. \#20). Although the single-output configurations were
preferred for processing of the INT stars, the uncertainty in the metallicity
derived from this configuration is similar to that achieved with the preferred
configuration, \#11, as visual comparison of Figure 9a with Figure 7a confirms. No
correlation between the errors in [Fe/H] and the color index were found, as
demonstrated in Figure 9b.

The quoted $\sigma_{\rm rms}$ value for the ${\rm [Fe/H]}_{\rm ANN}$ estimate
from the analyses of the standard stars included in the {\it testing} set,
which were not previously seen by the ANN,  is  0.32 dex.  This value
corresponds to configuration \#11, which was used to classify most of the
program stars, but the results for other configurations are very similar (see
Table 6).  The $\sigma_{\rm rms}$ values for $BV$  fall in the range
0.07--0.15, depending on the color itself.

There were 22 stars in the original INT sample with indices that were likely to
have been affected by an underlying emission component in the spectral lines
under consideration, as  simple visual inspection of their spectra suggests.
The metallicities derived for such stars using the ANN could be severely in
error, usually underestimated, and hence these stars were excluded.

\subsection{Comparison Between the ANN and Beers et al. (1999) Methods}

These two methods quote similar errors for the estimated $(B-V)_0$ colors of
the standard stars.  The ANN provides, using the KP and HP2 indices,
$\sigma_{\rm rms}$ values for $BV_{\rm ANN}$ on the order of 0.07 for $(B-V)_0
\le 0.8$, and on the order of 0.15 for $ (B-V)_0 > 0.8$.  The Beers et al.
(1999) calibration of the HP2 index yields errors for $BV_{\rm HP2}$ on the
order of 0.04 and 0.15 for these same color intervals, respectively.  At least
some of this error in predicted color, perhaps as much as half, may arise from
uncertainties in the reddening corrections applied to the standard stars.

Comparing the [Fe/H] estimates, the Beers et al. (1999) method exhibits a
nominal scatter of iron abundance estimates for the standard stars of 0.29 dex,
while the ANN reflects a scatter (for configuration \#11) of 0.32 dex.  These
error estimates {\it include} the contribution due to the error in the
abundances of the standards themselves, which are expected to be on the order
of 0.15--0.2 dex, so the intrinsic error of the techniques employed here is
$\sim$ 0.2 dex.

Clearly, the formal quoted errors are quite similar for both methods.  They
share the same standard stars, but the ANN uses only half of them for the
training sample.  Figure 10 shows the comparison of the results obtained using
both methods.  The standard deviation of the differences for the estimated
$(B-V)_0$ colors is 0.03 magnitudes, while the same quantity for the
differences in the predicted metallicities is 0.24 dex, well within the
requirements for the agreement with the quoted errors for each method.  There
is a low-order ``wiggle'' in the residuals of the comparison of abundances and
estimated colors obtained by the two methods, as can be seen in Figure 10, that
is presently of unknown origin.  

The ANN method is able to achieve  an accuracy similar to the Beers et al.
(1999) approach for estimates of $(B-V)_0$ and [Fe/H].  Its character is
completely empirical; the ANN is trained to reproduce the classification scheme
employed for the standard stars.  We expect that errors in the modeling of
stellar atmospheres will affect both the calibrations, but the former one
directly, for it is based on Kurucz's models and empirical corrections,  and
the latter one only indirectly for the standards having been analyzed with
classical model atmospheres.  The main advantage of the ANN estimator is of a
practical nature, as the effort required to tune a complex semi-empirical
procedure such as that employed by Beers et al. is generally greater than that
needed to properly train the ANNs.

\section{The Effective Yield of the INT Metal-Poor Star Follow-Up}
\label{sec4}

We have applied the trained ANNs to estimate [Fe/H] and $(B-V)_0$ for the 731
late-type stars in the INT follow-up. The normalized metallicity distribution
of the program stars, as derived from the ANN, is shown in Figure 11 (solid
histogram).  Comparison with the metallicity distribution of the metal-poor
candidates from the southern HK survey observed by BPS (dashed histogram)
indicates that the INT search has been far less effective in finding very
metal-poor stars than the southern program.  The peak of the metallicity
distribution for stars studied by BPS is centered around [Fe/H] $= -2.2$, while
the INT distribution is centered at approximately [Fe/H] $= -0.8$.  There are
several reasons for this difference.  Firstly, the identification of candidates
from the plates of the southern HK survey was,  intentionally, much more
selective than that applied to the plates of the northern hemisphere. In the
north, this choice was made in order to obtain larger samples of the
kinematically interesting metal-weak thick-disk stars.  Secondly, the southern
survey benefited from a partial photometric ``pre-filtering'' of
candidates based on $UBV$ photometry -- so that a greater fraction of the truly
metal-poor stars were observed spectroscopically, and the clear ``mistake''
stars, with colors either too blue or too red to be of interest, were
eliminated prior to spectroscopic follow-up.  No such photometric pre-filtering
step was taken for the INT program stars.  Furthermore, the regions of high
Galactic latitude (above $|b| = 60\deg$) were given higher priority in the
southern HK survey follow-up, but this was not so for the INT follow-up.
Figure 12 also shows that there also exists a tendency of the INT sample
towards brighter stars ($\langle$V(INT)$\rangle$=13.5 $\pm$ 0.8 mag, while
$\langle$V(BPS)$\rangle$=14.2 $\pm$ 0.6  mag), which would emphasize thick disk
stars relative to the generally fainter halo objects.

As a result of the above, a large proportion of the INT program stars of
intermediate metallicity are expected to belong to the thick-disk population,
while only the stars in the tail of the distribution at very low metallicity
are expected to be dominated by members of the Galactic halo population.
Figure 13, a comparison of the distribution of heliocentric radial velocities
for the INT sample (solid histogram) with the southern stars in the BPS sample
(dashed histogram), also bears this out.

The 731 stars analyzed in this work were identified from among the 1203
objects observed spectroscopically with the INT.  The survey effort has
identified 195 stars with [Fe/H] $\le -1.0$, 67 stars with [Fe/H] $\le -2.0$,
and 12 new stars with [Fe/H] $\le -3.0$.  Thus, using the definition of Beers
(2000), the {\it effective yield} (EY) of the INT spectroscopic follow-up for
stars below [Fe/H] $= -2.0$ is only on the order of 6\%.  Had a photometric
pre-filtering been done first, at a minimum the EY could have been boosted to
roughly 10\%, by elimination of the stars that are either too hot or too cool
to be of interest for this survey. If complete $UBV$ information were in hand
before the spectroscopic follow-up commenced, the EY would likely be on the
order of 20\%-30\%, thus commensurate with that found by other collaborations
that {\it did} perform a photometric pre-filtering (see Table 1 of Beers 1999).
Of course, the great advantage of proceeding with the spectroscopic follow-up
first is that the observations were conducted over only two observing seasons,
as opposed to many years of effort required for a complete photometric
pre-filtering, combined with later spectroscopy.

\section{Summary and Conclusions}
\label{sec5}

We have carried out a  search for metal-poor stars in the northern hemisphere
using the Isaac Newton Telescope.  Medium-resolution spectra of 1203 candidates
identified from the HK survey were obtained, from which 731 stars were selected
for further analysis.  Artificial Neural Networks were trained and tested with
previously available line indices for the large sample of standard stars
described by Beers et al. (1999), then were used to obtain estimates of [Fe/H]
and the intrinsic $(B-V)_0$ color index for the INT program stars.  This survey
effort has identified 195 stars with [Fe/H] $\le -1.0$, 67 stars with [Fe/H]
$\le -2.0$, and 12 new stars with [Fe/H] $\le -3.0$.  Although the EY of
metal-poor stars obtained was very low, the number of metal-poor stars
identified per observing hour was quite high, in a domain were multi-object
spectrographs on similar size telescopes cannot presently compete due to the
low areal density of targets, $\sim 2-4$ per square degree.

We conclude that ANNs are able to provide an accuracy of better than 0.3 dex in
the determination of stellar metallicity, and roughly 0.1 magnitudes in
estimates of broadband colors from intermediate-resolution spectra.  These
levels of accuracy are equivalent to those achieved by a much more elaborate
methodology that makes use of model atmospheres and empirical corrections.
ANNs can be trained rapidly, and can provide very fast classifications of
stellar spectra.  We emphasize that the procedure is {\it entirely} empirical,
the errors being driven mainly by the accuracy and consistency with which we
can provide metallicities for the {\it training} set of standard stars.  From
results previously obtained feeding the entire stellar spectrum to the ANNs (as
opposed to line indices) for determination of spectral classes  (Vieira \& Ponz
1995; Bailer-Jones 1997; Bailer-Jones et al.  1998), and recent tests on
samples spanning a large range of metallicities and temperatures (Qu et al.
1998; Snider et al.  2000), the application of the full-spectrum technique to
carry out multi-dimensional classification of stars appears very promising.

\bigskip
\bigskip

\acknowledgements

We sincerely acknowledge the efficient and professional help of the
staff at the Observatorio del Roque de los Muchachos, especially
Palmira Arenaz, Marco Azzaro, Miriam Centuri\'on,  Carlos Mart\'{\i}n,
Neil O'Mahony, and Juerg Rey in the operation of the INT and the IDS
spectrograph. David Lambert is thanked for his participation in one of
the observing runs and for interesting discussions. We are  indebted to the referee  for suggestions that helped to clarify the presentation.

This research has made use of the SIMBAD database, operated at CDS, Strasbourg
(France), the  NASA's ADS Abstract Service, and has been partially funded by
the Spanish DGES under projects  PB95-1132-C02-01 and PB98-0531-C02-02.  This work also
received partial support from NATO grant 950875, and extensions.  TCB
acknowledges partial support from grant AST 95-29454 awarded by the US National
Science Foundation.  SR acknowledges partial support from the Brazilian
agencies CNPq (grant 200068/95-4) and FAPESP (98/02706-6).

\clearpage



\begin{figure}[ht!]
\centering
\includegraphics[width=10.cm,angle=90]{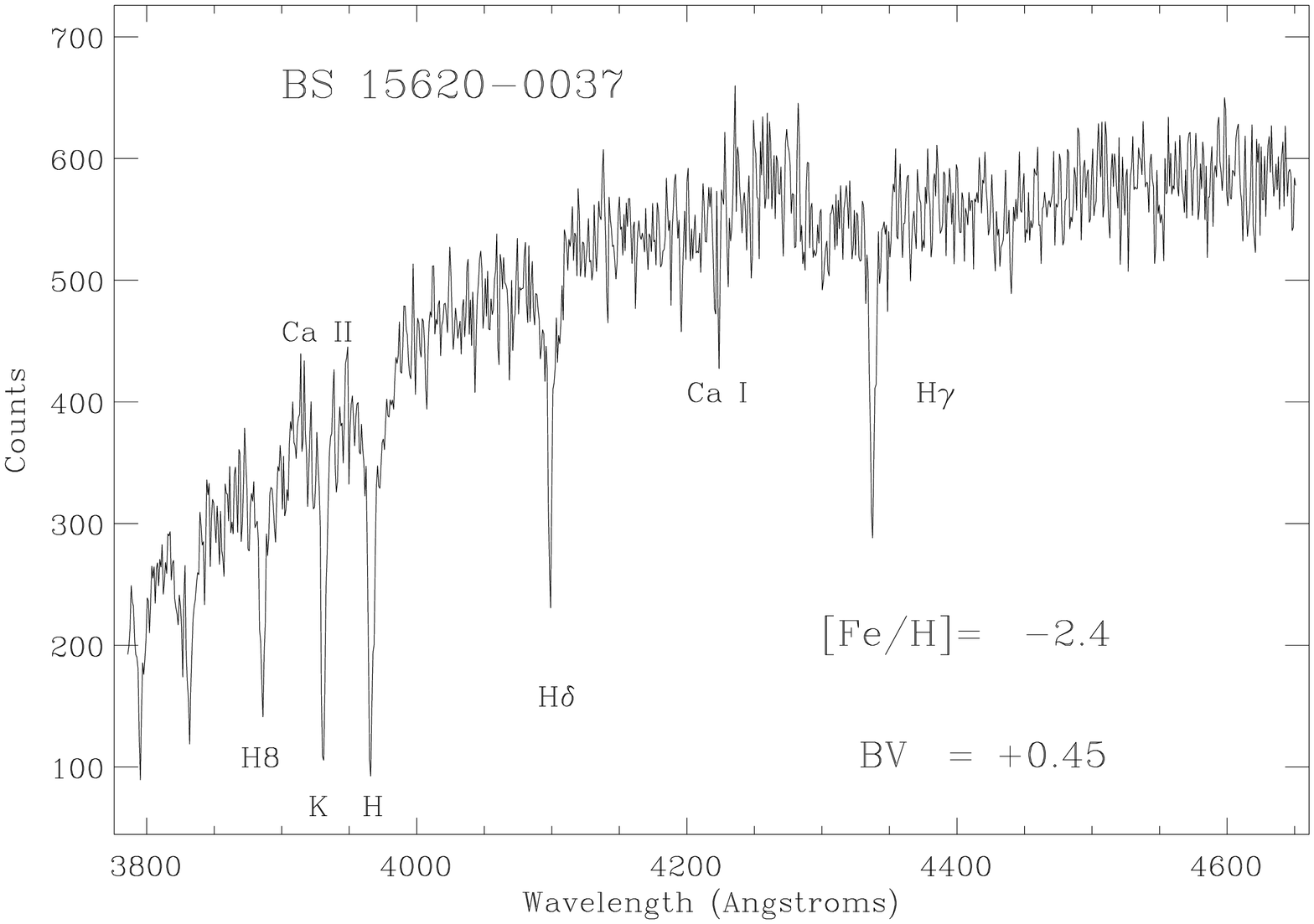}  
\protect\caption[]{Typical spectrum of a candidate metal-poor star from the INT
follow-up campaign.  The prominent spectral features are labelled, along with
the derived estimate of metallicity and broadband color..
\label{fig1}}
\end{figure}

\begin{figure}[ht!]
\centering
\includegraphics[width=10.cm,angle=90]{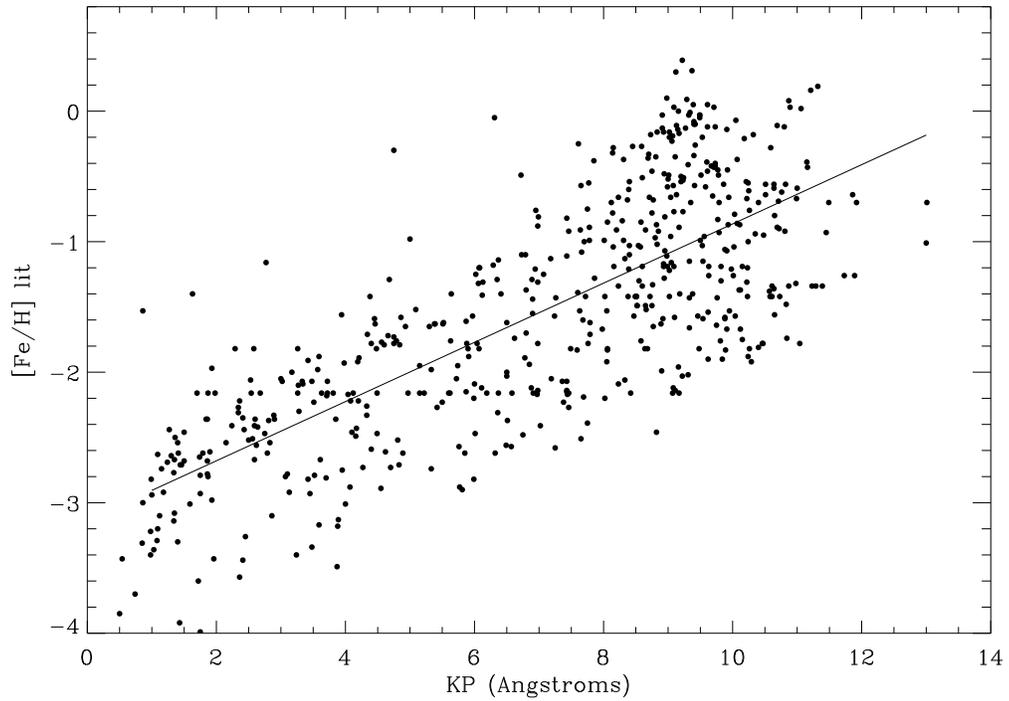}  
\protect\caption[]{High-resolution metallicity determination, [Fe/H]$_{\rm lit}$, versus the
KP index for the standard stars of Beers et al. (1999) used in the training of
the ANN's.  The solid line is a least-square linear fit.  The scatter arises
primarily from the range of temperatures of the stars at a given abundance.}
\end{figure}

\begin{figure}[ht!]
\centering
\includegraphics[width=10.cm,angle=90]{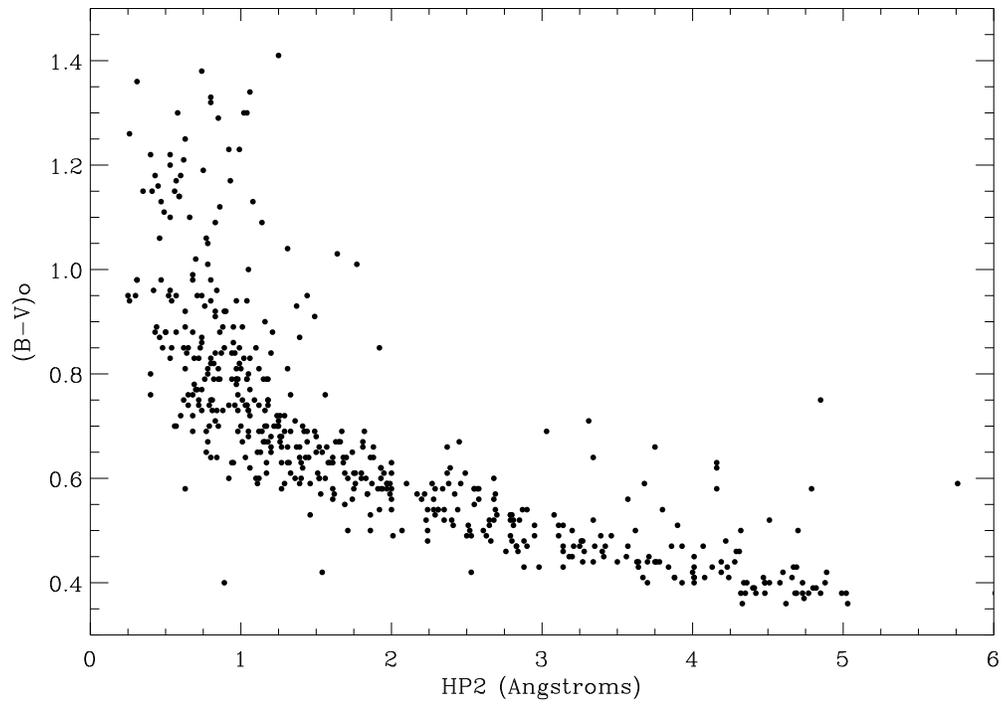}  
\protect\caption[]{De-reddened Johnson $(B-V)_0$ color versus the HP2 index for the
Beers et al. (1999) standard stars, over the range of the HP2 index
for stars considered in this work.}
\end{figure}

\begin{figure}[ht!]
\centering
\includegraphics[width=10.cm,angle=0]{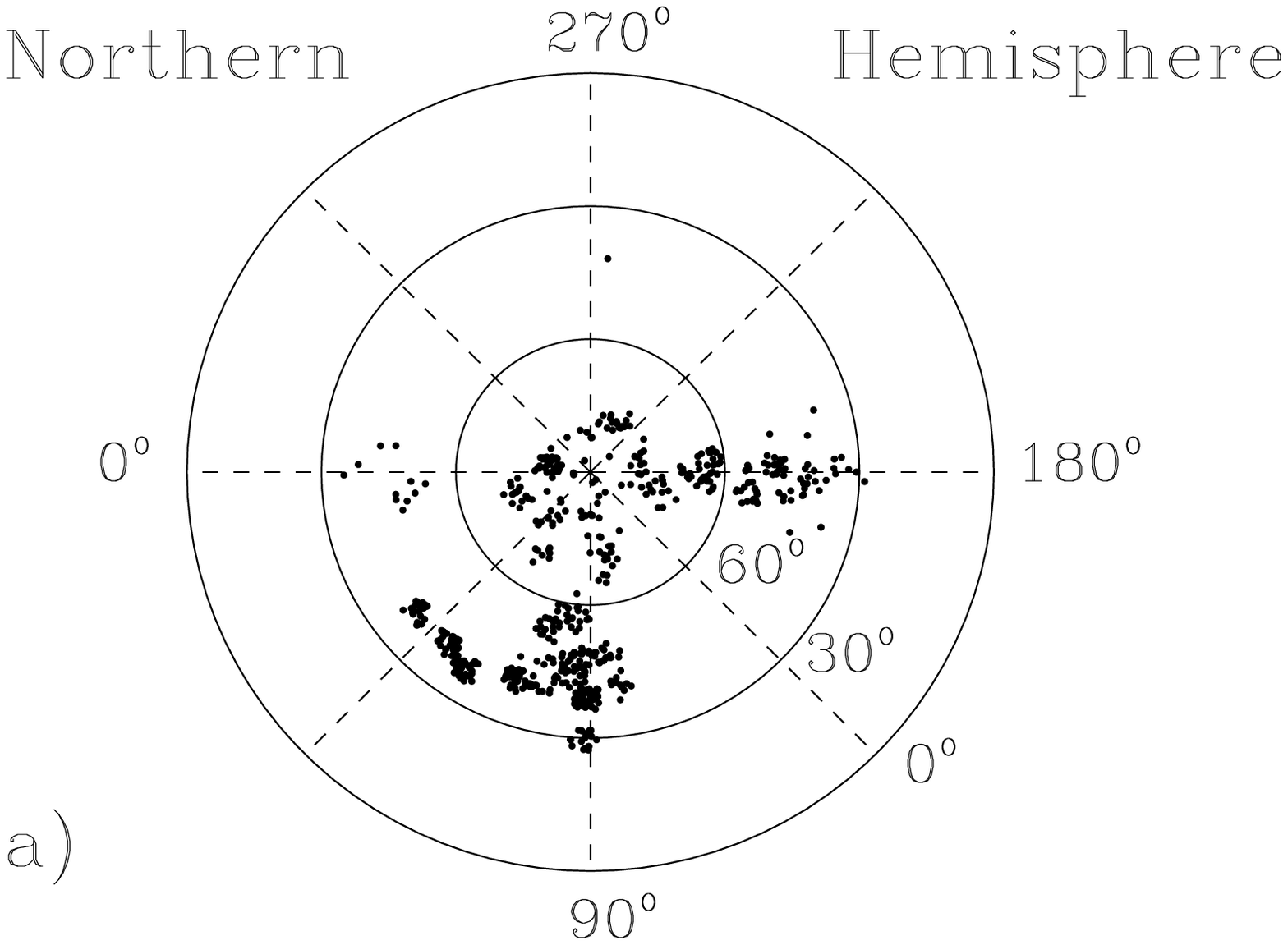}  
\includegraphics[width=10.cm,angle=0]{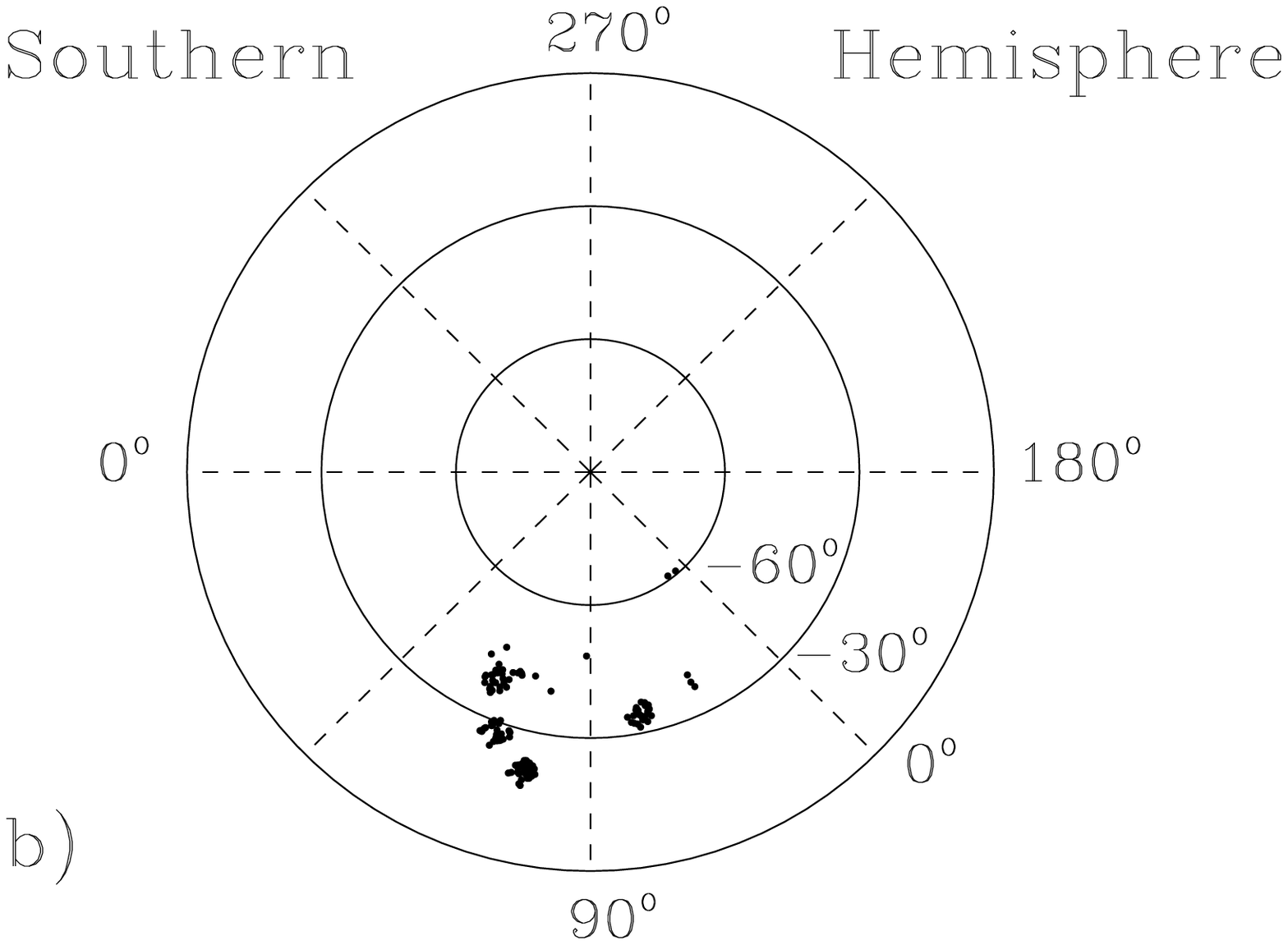}  
\protect\caption[]{Galactic coordinates of the INT program stars. The Galactic
longitude, in degrees, corresponds to the angular coordinate in the polar plot,
while the radial coordinate represents the Galactic latitude, also in degrees.
(a) Northern Hemisphere, (b) Southern Hemisphere.}
\end{figure}

\begin{figure}[ht!]
\centering
\includegraphics[width=10.cm,angle=90]{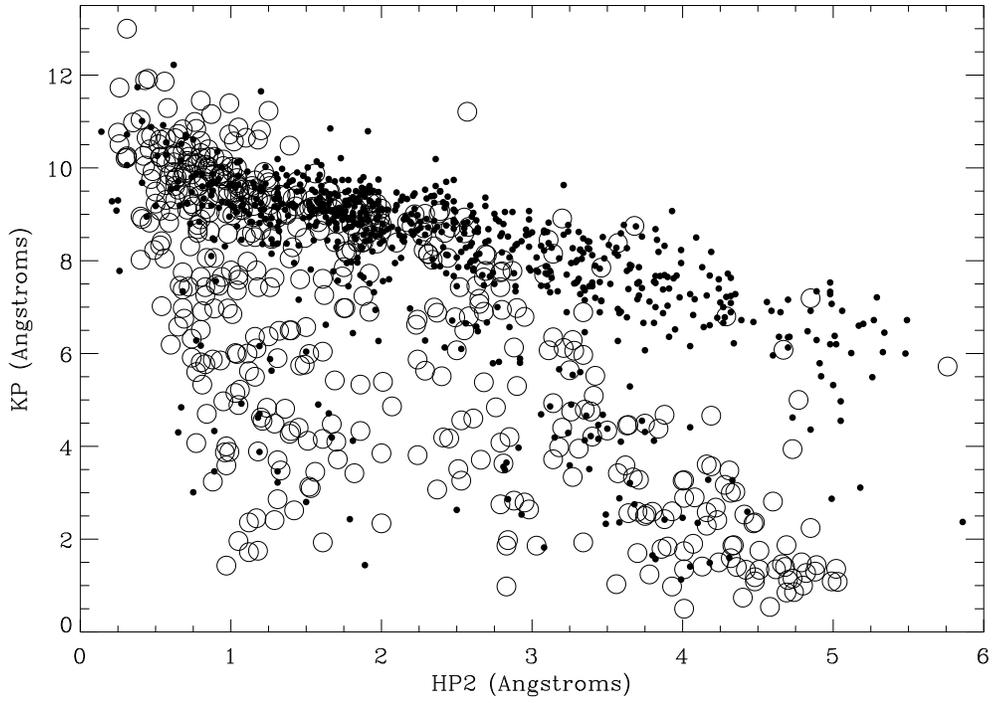}  
\protect\caption[]{Positions of the standard stars (open circles), and the program
stars (dots) in the KP--HP2 plane.}
\end{figure}

\begin{figure}[ht!]
\centering
\includegraphics[width=10.cm,angle=90]{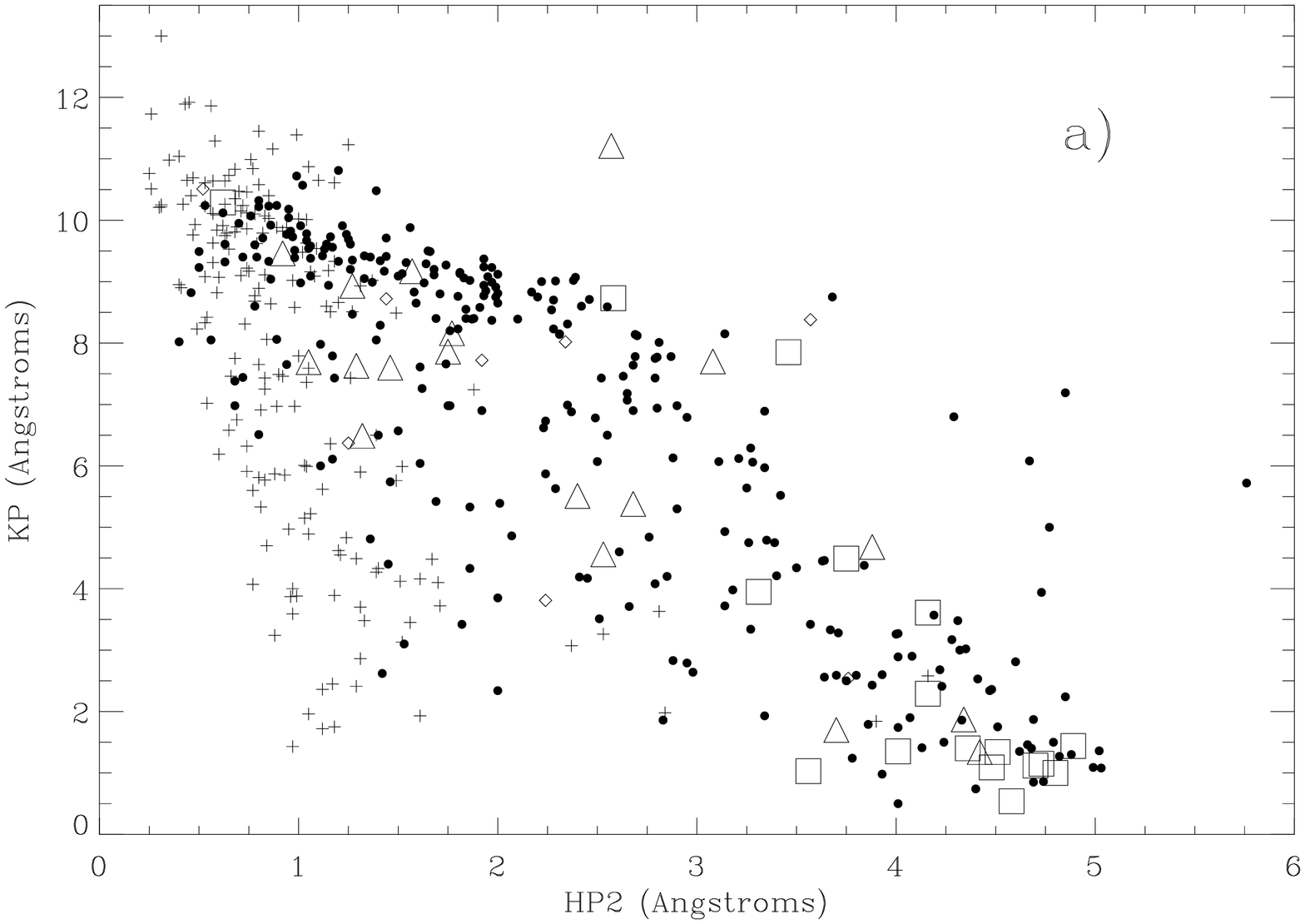}  
\includegraphics[width=10.cm,angle=90]{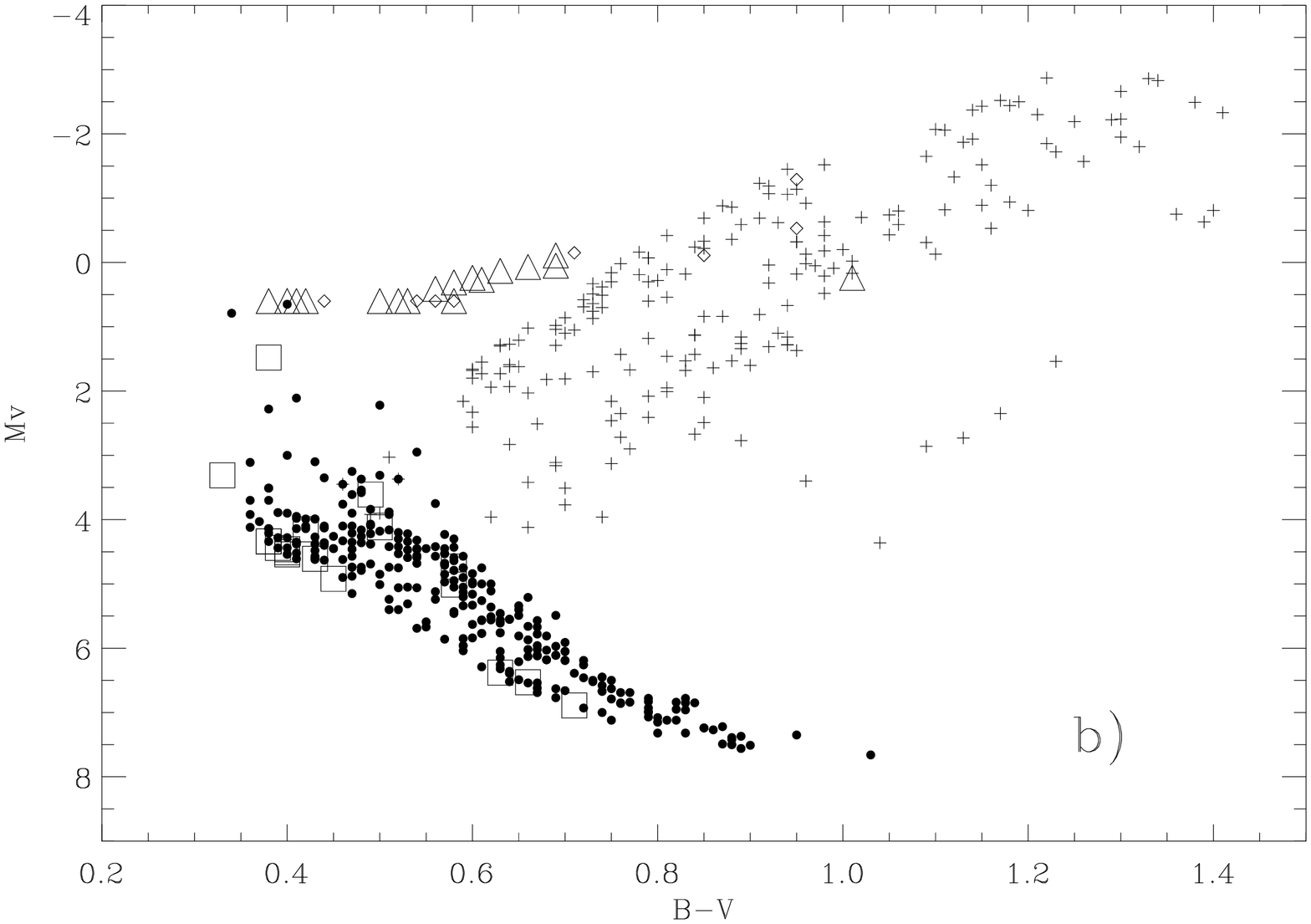}  
\protect\caption[]{(a) Positions of the Beers et al. (1999) standard stars in the
KP--HP2 plane. The codes correspond to stars of different types, as adopted by
Beers et al.:  Dwarfs -- dots; Giants -- crosses; Horizontal-branch stars --
triangles; Asymptotic giant-branch stars -- rhombi; Main-sequence turn-off
stars -- squares. (b) Positions of the same standards in the
Hertzsprung--Russell diagram.}
\end{figure}

\begin{figure}[ht!]
\centering
\includegraphics[width=10.cm,angle=90]{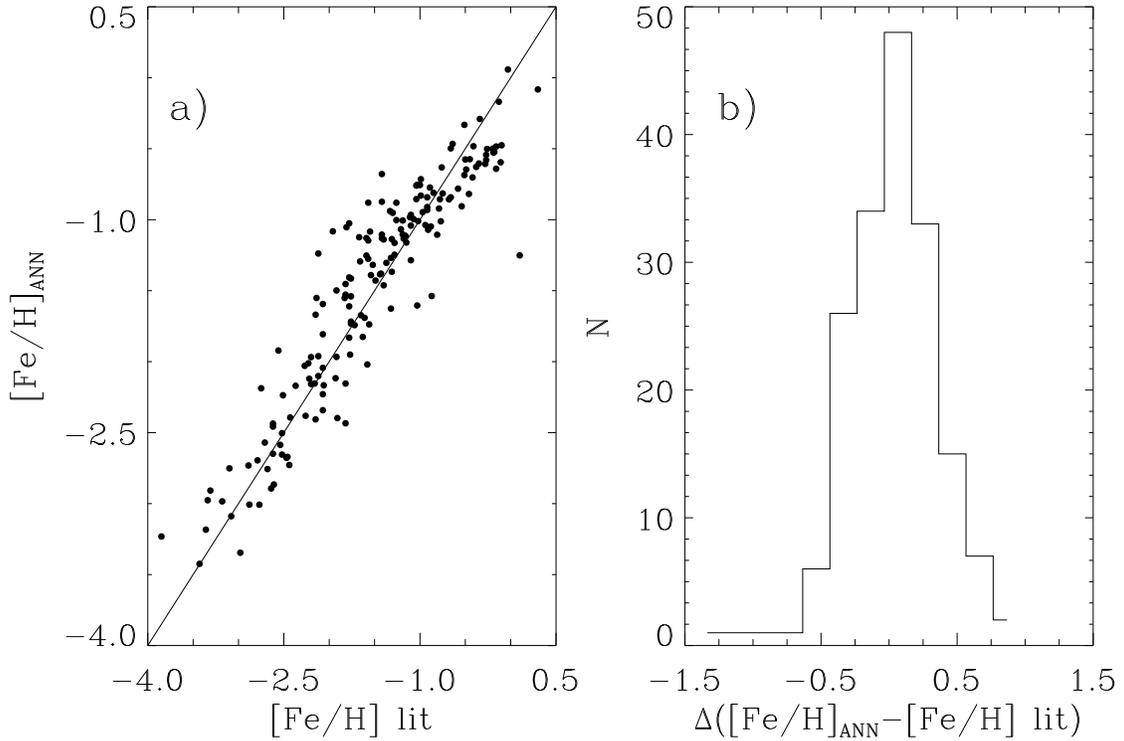}  
\protect\caption[]{(a) Metallicities of the {\it testing}  standard stars, as derived
from the  ANN (single output neuron) configuration \#11, [Fe/H]$_{\rm ANN}$,
compared with published values, [Fe/H]$_{\rm lit}$, derived  from
high-resolution spectroscopic analyses.  The solid line is the one-to-one
relation. (b) Distribution of the differences between the [Fe/H]s predicted by
ANN configuration \#11 and those published in high-resolution spectroscopic
studies. The {\it testing} sample of standards is classified by the ANN giving
a $\sigma_{\rm rms} = 0.32$ dex in [Fe/H], while the mean of the difference
between the [Fe/H]$_{\rm ANN}$ and [Fe/H]$_{\rm lit}$ values is  $+$0.04 dex.}
\end{figure}

\begin{figure}[ht!]
\centering
\includegraphics[width=10.cm,angle=90]{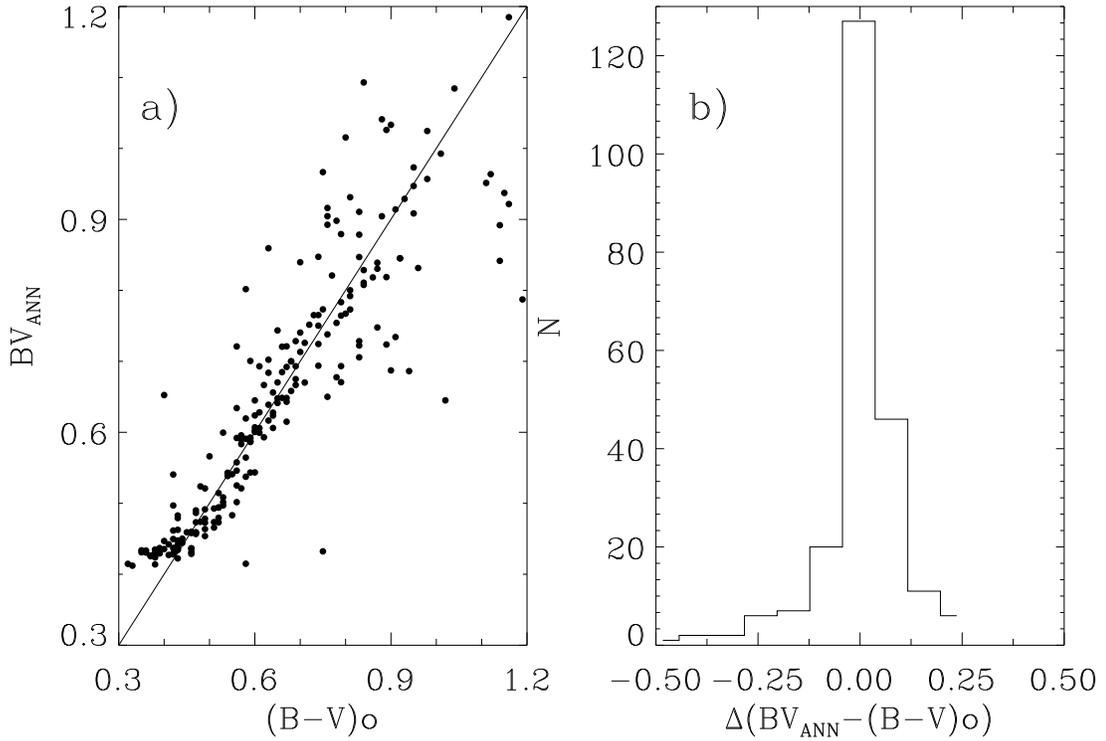}  
\protect\caption[]{(a) Predicted $(B-V)_0$ color, $BV$, for the {\it testing} sample
of standard stars, as derived from the ANN (single output neuron) configuration
\#4, compared with published photometric values.  The solid line is the
one-to-one relation.  (b) Distribution of the differences between $BV$
predicted by the ANN and the de-reddened measured photometric values.  The
expected error in $BV$ depends strongly on the color itself, the errors being
much larger for cooler stars than for the hotter stars. The error
distribution is clearly non-Gaussian.}
\end{figure}

\begin{figure}[ht!]
\centering
\includegraphics[width=10.cm,angle=90]{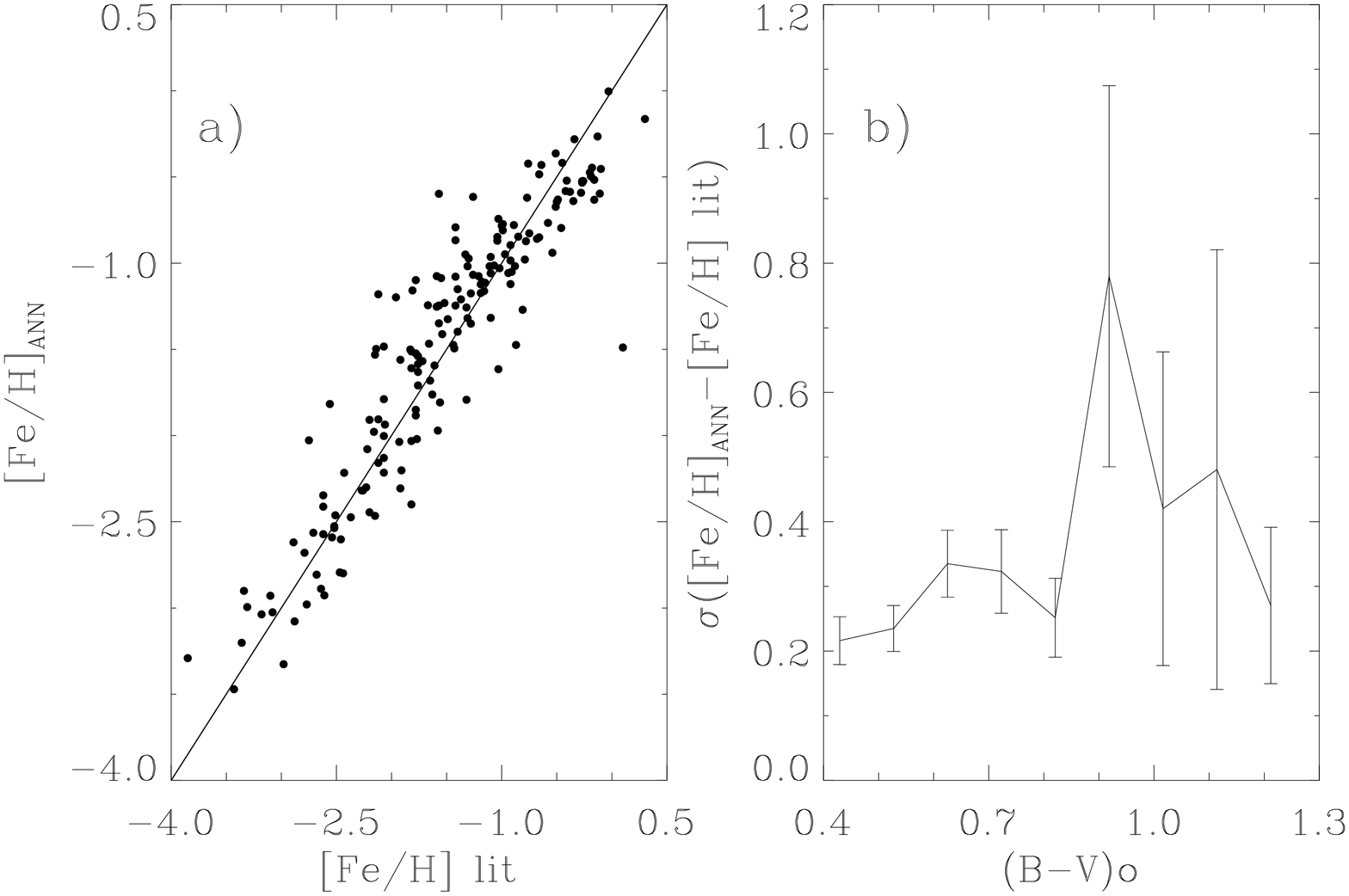}  
\protect\caption[]{ (a) Metallicities of the {\it testing} sample of standard stars,
as derived from the ANN (two output neurons) configuration \#20, [Fe/H]$_{\rm
ANN}$, compared with published values, [Fe/H]$_{\rm lit}$, derived from
high-resolution spectroscopic analyses.  The solid line is the one-to-one
relation. (b) Distribution of the standard deviation of differences between the
[Fe/H]'s predicted by ANN configuration \#20 and those published in
high-resolution spectroscopic studies, as a function of the $(B-V)_0$ color.
The error bars scale inversely proportional to the square root of the number of
stars. No correlation  between the errors in the ANN metallicity determination
and the color index is apparent.}
\end{figure}

\begin{figure}[ht!]
\centering
\includegraphics[width=10.cm,angle=90]{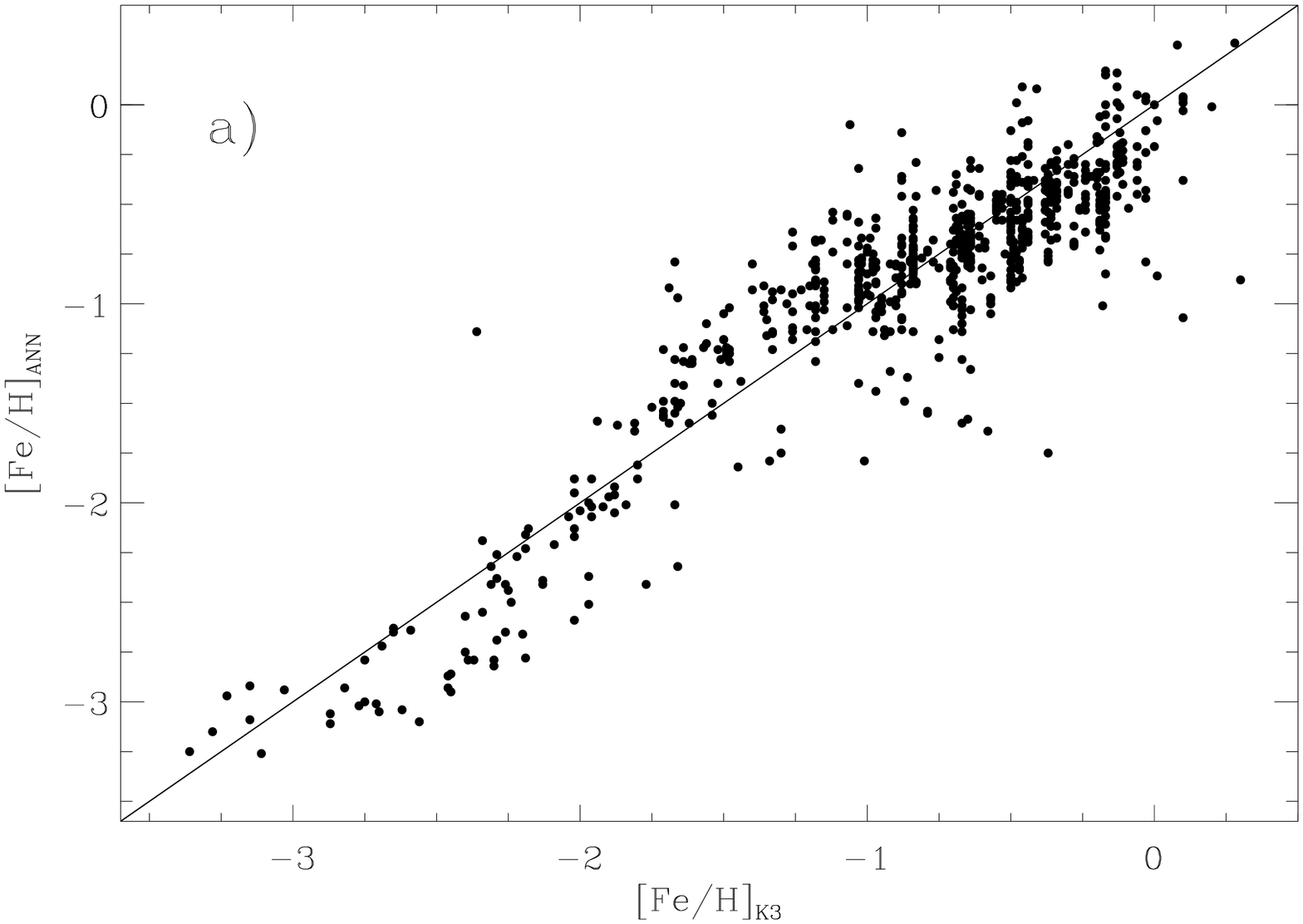}  
\includegraphics[width=10.cm,angle=90]{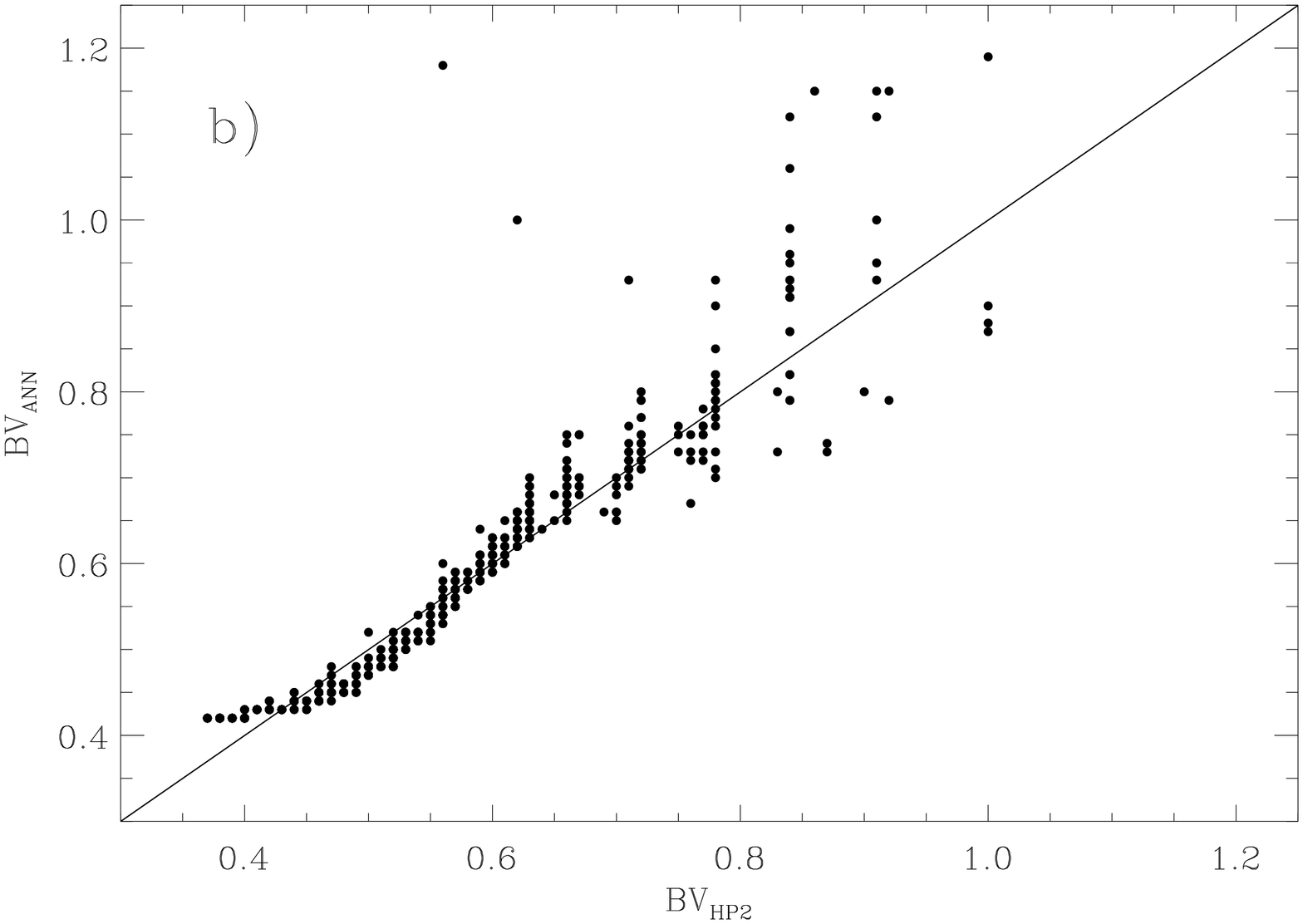}  
\protect\caption[]{(a) Comparison between the metallicities estimated for the standard
stars derived from the ANN approach, [Fe/H]$_{\rm ANN}$, and the estimate
[Fe/H]$_{\rm K3}$ from Beers et al.  The solid line is the one-to-one relation.
(b) Comparison between the estimated $(B-V)_0$ for the set of Beers et al.
(1999) standards derived using the ANN approach, $BV_{\rm ANN}$, and the HP2
calibration of Beers et al., $BV_{\rm HP2}$.}
\end{figure}

\begin{figure}[ht!]
\centering
\includegraphics[width=10.cm,angle=90]{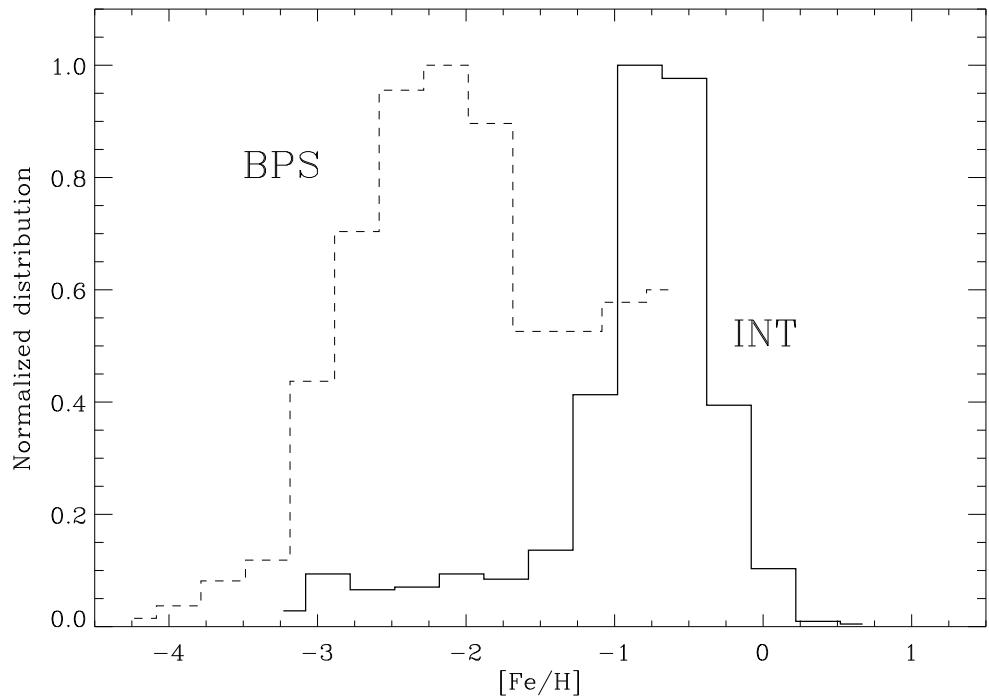}  
\protect\caption[]{Metallicity distributions of the INT program sample (solid
histogram) and the stars from BPS (dashed histogram). The BPS follow-up was
clearly much more successful in the detection of extremely metal-poor stars
(but see discussion in text). }
\end{figure}

\begin{figure}[ht!]
\centering
\includegraphics[width=10.cm,angle=90]{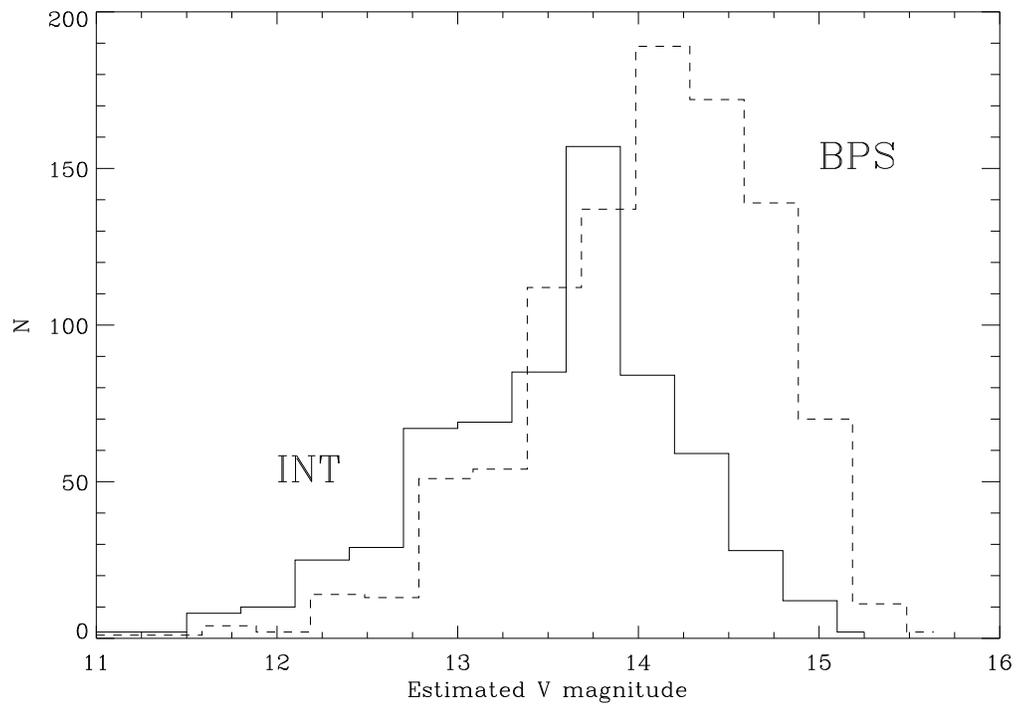}  
\protect\caption[]{Distribution of $V$ magnitudes for the INT program sample (solid
histogram) and the BPS follow-up (dashed histogram). The INT stars are
generally brighter than the BPS stars.}
\end{figure}

\begin{figure}[ht!]
\centering
\includegraphics[width=10.cm,angle=90]{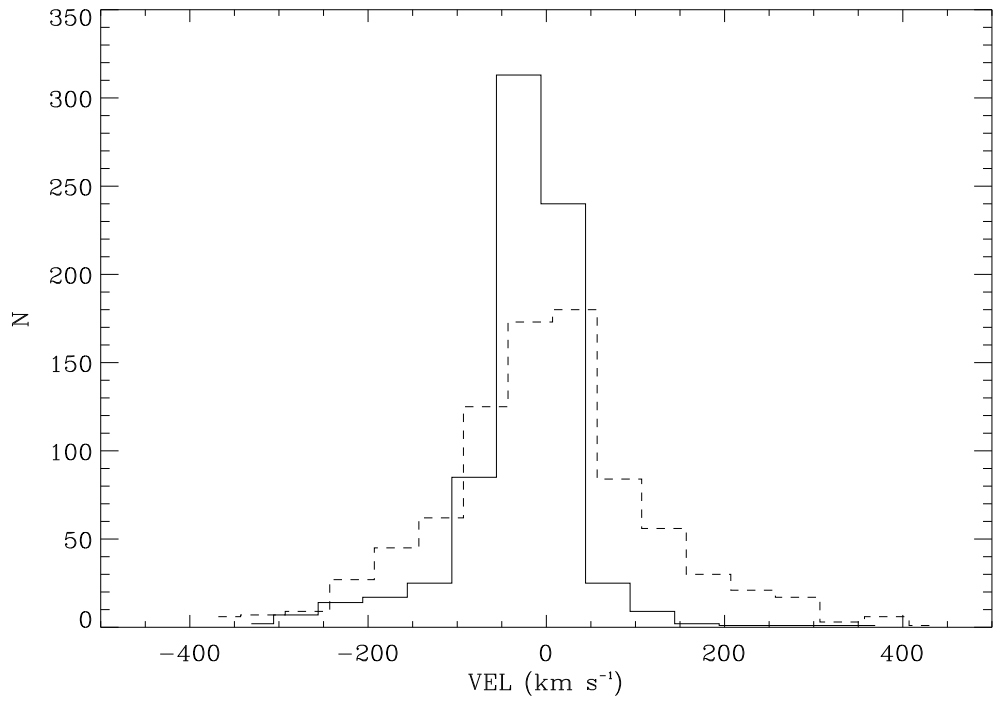}  
\protect\caption[]{ Radial-velocity distributions for the INT program sample
(solid histogram) and the BPS follow-up (dashed histogram).  Note the much
broader tails in the line-of-sight velocities for the BPS sample, indicative of
a greater fraction of halo stars.}
\end{figure}

\end{document}